\documentclass[aps,prc,twocolumn,superscriptaddress,showkeys,nofootinbib]{revtex4-1}

\makeatletter
\renewcommand\subsection{\@startsection{subsection}{2}{\z@}
  {-2.0ex\@plus -.6ex \@minus -.2ex}
  {0.8ex \@plus .2ex}
  {\normalfont\normalsize\itshape}}
\makeatother

\usepackage{amsmath,amssymb}

\begin{document}

\title{Comment on ``Non-Monotonicity of Transverse-Momentum Correlations \\
in Au+Au Collisions at RHIC''}

\author{Roy A. Lacey}
\affiliation{Depts. of Chemistry and Physics, Stony Brook University, Stony Brook, NY 11794, USA}

\begin{abstract}
Recent measurements of two-particle transverse-momentum correlations,
$C_{p_T}$, in Au+Au collisions have revealed a statistically significant
non-monotonic beam-energy dependence over the range
$\sqrt{s_{NN}}=3.0$--$7.7$~GeV that has been interpreted as potentially
indicative of critical phenomena associated with a critical end point
(CEP) in the QCD phase diagram. This interpretation is assessed in light
of the absence of a controlled framework establishing $C_{p_T}$ as a
quantitative proxy for the underlying critical response. The implications
of this limitation are examined by considering the expected finite-size
and finite-time modification of critical fluctuations, substantial
non-critical dynamical contributions, the lack of demonstrated critical
scaling, and the sizeable discrepancy between the CEP region inferred
from the observed $C_{p_T}$ non-monotonicity and that constrained by
susceptibility-based measurements and finite-size scaling analyses.
Taken together, these considerations indicate that the observed
$C_{p_T}$ non-monotonicity does not, by itself, establish a critical
origin nor provide reliable quantitative constraints on the existence
or location of a possible CEP.
\end{abstract}

\maketitle

Event-by-event fluctuations in relativistic heavy-ion collisions are widely used to probe the QCD phase diagram and search for a possible critical end point (CEP) \cite{Stephanov:1998dy,Stephanov:1999zu}. In this context, the STAR Collaboration has reported measurements of higher-order cumulants of conserved charges \cite{STAR:2014egu,STAR:2020tga,STAR:2025zdq} as well as two-particle transverse-momentum correlations, $C_{p_T}$ \cite{STAR:2008szd}. Recent STAR BES-II measurements of $C_{p_T}$ in Au+Au collisions spanning the high-$\mu_B$ region of the QCD phase diagram ($\sqrt{s_{NN}}=3.0$--$7.7$~GeV, corresponding to $\mu_B\sim760$--$400$~MeV) exhibit a statistically significant non-monotonic beam-energy dependence that has been interpreted as potentially indicative of a CEP \cite{STAR:2026vjv}.

The central question is whether this non-monotonicity can be quantitatively connected to the equilibrium critical behavior associated with a CEP. Establishing such an interpretation requires a controlled framework that connects the measured observable, or an appropriate proxy, to the underlying critical response. This Comment examines whether such a framework has been established for $C_{p_T}$, with particular emphasis on consistency with susceptibility-based constraints from higher-order cumulants of conserved charges and finite-size scaling analyses.

The observable $C_{p_T}$ quantifies event-by-event fluctuations of the mean transverse momentum. Although, in an equilibrated system of fixed volume, fluctuations of $\langle p_T \rangle$ can be related to temperature fluctuations and the heat capacity \cite{Stephanov:2008qz}, this connection is not direct in heavy-ion collisions. Owing to rapid expansion, strong collective flow, and finite size and lifetime, the measured $C_{p_T}$ reflects the combined influence of equilibrium and non-equilibrium dynamics, including fluctuations of radial flow, hadronic rescattering, baryon transport, resonance decays, diffusion, and changes in particle composition \cite{Ollitrault:1992bk,Teaney:2003kp,Heinz:2013th}. Thus, while $C_{p_T}$ may retain sensitivity to equilibrium thermodynamic fluctuations, interpreting it as a quantitative proxy for such fluctuations requires a controlled framework that accounts for these additional
dynamical contributions.

This distinction is fundamental because theoretical predictions for the CEP are based on the equilibrium thermodynamic equation of state, whereas the measured $C_{p_T}$ reflects the evolution of a finite, rapidly expanding system. Equilibrium critical behavior associated with a CEP is encoded in QCD thermodynamic susceptibilities, making observables that directly access these susceptibilities among the most direct experimental probes currently available for investigating such behavior. An experimentally accessible observable need not itself be a thermodynamic susceptibility to provide information on critical behavior; it may instead serve as a proxy for the underlying critical response, provided that its connection to that response is established within a controlled framework. Such a framework must quantitatively relate the observable to the underlying critical response and establish the expected behavior under changes in the relevant control variables and system size. No well-established framework presently demonstrates such a connection or critical scaling for $C_{p_T}$. Consequently, interpreting $C_{p_T}$ as a quantitative proxy for temperature fluctuations or as evidence for equilibrium critical behavior remains model dependent at best.

Even with such a framework, finite-size and finite-time effects impose a further limitation. Near a CEP, the growth of equilibrium critical fluctuations is constrained: the correlation length remains bounded, and critical slowing down limits the development of equilibrium correlations \cite{Berdnikov:1999ph}. Subsequent expansion, diffusion, and hadronic rescattering further modify these developing correlations. Taken together, these effects suppress the observable critical signal, broaden its beam-energy dependence, and can shift the apparent location of its extremum relative to the underlying equilibrium behavior. Moreover, for the three-dimensional Ising universality class expected for the QCD CEP \cite{Stephanov:1998dy,Stephanov:1999zu}, the small heat-capacity critical exponent implies a comparatively weak singular response in temperature-related fluctuations relative to higher-order susceptibility observables. Thus, if the critical contribution to $C_{p_T}$ is mediated primarily through temperature or heat-capacity fluctuations, it is expected to be comparatively modest even before finite-size, finite-time, and dynamical effects are taken into account. These effects would further attenuate and broaden such a contribution and could shift the location of its extremum. Consequently, the pronounced magnitude and relatively localized character of the observed $C_{p_T}$ non-monotonicity warrant caution in assigning it
a critical origin. Moreover, without a controlled quantitative mapping,
the beam energy of the observed extremum cannot be identified directly
with the location of the underlying thermodynamic CEP.

The distinction between a statistically significant deviation from an assumed baseline and statistically significant evidence for critical behavior is also important. The reported significance of the observed non-monotonicity is quantified relative to a baseline comprising a $1/\sqrt{N_{\mathrm{part}}}$ scaling, motivated by the independent-source picture in which particles are emitted from uncorrelated sources, together with a low-order polynomial representing the assumed monotonic beam-energy dependence. The independent-source assumption is progressively violated by collective expansion and hadronic dynamics that generate correlated particle production, while the polynomial reference is not uniquely prescribed by theory. The reported significance therefore quantifies the incompatibility of the measurements with these assumed baselines, but does not by itself establish the physical origin of the deviation. In particular, without a controlled framework linking the observed deviation to the underlying critical response, its statistical significance cannot be interpreted as a corresponding significance for critical behavior.

The physical origin of such a deviation must therefore be established
independently. Across the BES energy range, the collision dynamics
introduce multiple non-critical contributions to $C_{p_T}$ whose relative
importance evolves strongly with beam energy. As $\sqrt{s_{NN}}$ decreases,
increasing baryon stopping and net-baryon density drive a transition from
a meson-dominated regime toward a baryon-rich system. Hadronic rescattering,
baryon transport, changes in particle composition, antibaryon annihilation,
and the evolution of the effective equation of state provide additional
energy-dependent contributions to the measured correlations
\cite{Bass:1998ca,Bleicher:1999xi,Lacey:2024bcm}. Recent calculations
demonstrate explicitly that the changing mixture of mesonic and baryonic
contributions can generate a non-monotonic beam-energy dependence of
transverse-momentum correlations without invoking criticality
\cite{Reichert:2026ztj}. Together, these coupled dynamical processes
provide multiple mechanisms through which non-monotonic behavior can
emerge from the evolving collision dynamics.

The observed centrality dependence provides a direct experimental manifestation of this dynamical complexity. While a pronounced non-monotonic structure is reported in the most central (0--5\%) collisions, corresponding to a significance of approximately $5\sigma$, a substantially reduced effect is observed in mid-central (30--40\%) events, where the reported significance is $\sim2\sigma$, despite the overall magnitude of $C_{p_T}$ being approximately a factor of two larger than in the 0--5\% collisions \cite{STAR:2026vjv}. This striking contrast demonstrates that the reported non-monotonic structure does not simply track the overall magnitude of the measured fluctuations, but instead points to a strong centrality dependence in the relative contributions of the physical mechanisms governing the measured correlations. Although this behavior does not exclude a critical contribution, it demonstrates that the mechanisms contributing to $C_{p_T}$ evolve substantially with centrality and therefore further complicate a unique attribution of the observed non-monotonicity to critical behavior.

Comparisons with available transport and Boltzmann--Langevin calculations further underscore the present limitations in quantitatively disentangling these contributions. No existing framework has yet demonstrated a comprehensive quantitative description of the principal non-critical processes across the full range of beam energies and collision centralities. Developing such a framework remains an active area of research, with recent efforts emphasizing the coupled roles of baryon transport, hadronic chemistry, rescattering, and the effective equation of state governing the collision dynamics across the BES energy range \cite{Lacey:2024bcm}. Consequently, the inability of available models to reproduce the pronounced non-monotonic structure observed in the most central collisions does not, by itself, discriminate between a genuinely critical contribution and limitations in the current quantitative description of the underlying non-critical dynamics. Model failure therefore cannot presently be regarded as independent evidence for a critical origin of the observed structure.

It is therefore natural to compare $C_{p_T}$ with observables that either
probe QCD thermodynamic susceptibilities more directly or serve as
experimentally accessible proxies for the underlying critical response
within a controlled scaling framework. Higher-order cumulants of conserved
charges provide the most direct experimental probes of these
susceptibilities, which encode the equilibrium critical behavior associated
with a CEP \cite{Stephanov:2008qz,STAR:2014egu,STAR:2020tga}. Like
$C_{p_T}$, these observables are also affected by finite-size, finite-time,
and dynamical effects. However, finite-size scaling provides a controlled
framework for identifying and quantitatively characterizing the underlying
critical behavior despite these limitations
\cite{Lacey:2015aza,Lacey:2024mnv}. In particular, finite-size scaling analyses of multiple independent higher-order cumulant ratios provide self-consistent constraints on a
common CEP region \cite{Lacey:2024mnv}. These results presently provide
the strongest quantitative experimental constraints on the QCD critical
end point.

Importantly, an experimental observable need not itself be a thermodynamic
susceptibility to provide quantitative information on critical behavior.
Finite-size scaling has, for example, been applied to the two-pion
interferometry observable $(R_{\rm out}^2-R_{\rm side}^2)$ as an
experimentally accessible proxy related to the compressibility of the
medium. The characteristic system-size dependence of its non-monotonic
excitation functions was used to extract both a CEP location and critical
exponents consistent with the three-dimensional Ising universality class
\cite{Lacey:2015aza}. More recently, finite-size scaling analyses of
multiple independent higher-order cumulant ratios, using critical exponents
fixed to their three-dimensional Ising values, yield robust scaling
functions and a compatible CEP region \cite{Lacey:2024mnv}. The consistency
of the inferred critical behavior across these distinct observables and
scaling strategies provides an important cross-check of both the
universality class and the CEP region.

These examples illustrate that non-monotonicity can motivate a critical
interpretation, but the quantitative basis for that interpretation is
provided by the additional demonstration of the expected finite-size
scaling behavior. More generally, the essential requirement is not that
the measured observable itself be a thermodynamic susceptibility, but that
its role as a probe or proxy for the underlying critical response be
established and tested within a controlled scaling framework. No analogous
scaling behavior has yet been demonstrated for $C_{p_T}$. Thus, the mere
appearance of a localized non-monotonic feature does not establish that
$C_{p_T}$ is a reliable proxy for critical behavior, and a visible
non-monotonicity is neither a necessary nor a sufficient basis for
establishing a critical interpretation.

In this context, because these observables and experimentally accessible
proxies probe the same underlying QCD phase structure in Au+Au collisions,
any interpretation of the observed beam-energy dependence of $C_{p_T}$ in
terms of critical behavior should ultimately be reconcilable with the CEP
constraints obtained from susceptibility-based observables, appropriate
proxies, and controlled scaling analyses. Finite-size effects can shift the
apparent location of a critical feature in the unscaled finite-system data
relative to the thermodynamic CEP. Finite-size scaling exploits the
characteristic system-size dependence of this shift to constrain the CEP
location in the thermodynamic limit \cite{Lacey:2015aza,Lacey:2024mnv}.
Thus, a modest displacement between the CEP region indicated by the
background-subtracted higher-order cumulant measurements \cite{STAR:2014egu,STAR:2020tga,STAR:2025zdq} and that obtained from finite-size scaling \cite{Lacey:2024mnv} is not unexpected.

The situation for $C_{p_T}$ is markedly different. Associating the location
of its observed non-monotonicity directly with a CEP places the inferred CEP
well outside the region indicated by the background-subtracted STAR
measurements of higher-order cumulants of conserved charges
\cite{STAR:2014egu,STAR:2020tga,STAR:2025zdq}, and even farther from the
thermodynamic CEP region obtained through controlled finite-size scaling
analyses \cite{Lacey:2015aza,Lacey:2024mnv}. This sizeable discrepancy is
far greater than the modest displacement between the finite-system
cumulant indication and the corresponding finite-size-scaled CEP region,
and therefore cannot simply be regarded as an expected finite-size shift.
Moreover, no controlled finite-size scaling or other quantitative framework
has established how the location of the observed $C_{p_T}$ extremum should
map onto the underlying thermodynamic CEP. The discrepancy therefore
reinforces the central limitation of the proposed interpretation: the
location of the observed $C_{p_T}$ non-monotonicity cannot presently be
used to provide a reliable quantitative constraint on the location of the
CEP.

Taken together, the methodological and phenomenological considerations
discussed here indicate that the observed non-monotonic beam-energy
dependence of $C_{p_T}$ does not, by itself, establish a critical origin
or provide reliable quantitative constraints on the QCD critical end
point. Higher-order cumulants of conserved charges provide direct
experimental access to the thermodynamic susceptibilities that encode
equilibrium critical behavior, while experimentally accessible proxies
can also provide quantitative critical information when their connection
to the underlying critical response is established within a controlled
scaling framework. Finite-size scaling analyses of both higher-order cumulant ratios and
a compressibility-related proxy demonstrate this approach, yielding
critical behavior consistent with the three-dimensional Ising
universality class and compatible CEP regions
\cite{Lacey:2015aza,Lacey:2024mnv}. By contrast, no comparable framework
has yet established $C_{p_T}$ as a controlled proxy for the underlying
critical response or demonstrated how the location of its observed
non-monotonicity maps onto the thermodynamic CEP. The sizeable discrepancy 
between the CEP region inferred from the $C_{p_T}$ extremum and that indicated 
by the background-subtracted STAR higher-order cumulant measurements, and even more strongly by finite-size scaling, further underscores this limitation. 
Until such a framework is established, $C_{p_T}$ cannot presently
provide reliable quantitative constraints on the QCD critical end point.
%Establishing such a framework is therefore essential before transverse-momentum 
%correlations can be used to draw quantitative conclusions about the existence or 
%location of the QCD critical end point.

%
\bibliography{Comments_pT_fluctations-refs}

\end{document}